\pdfoutput=1
\PassOptionsToPackage{bookmarks=false,colorlinks=true,linkcolor=blue,urlcolor=cyan,filecolor=magenta,citecolor=red,pdfstartview=FitV,hyperfootnotes=false,pdftitle={Carroll Hawking effect},pdfauthor={Ankit Aggarwal, Florian Ecker, Daniel Grumiller, Dmitri Vassilevich},pdfsubject={Carroll Hawking effect},pdfkeywords={Carroll Hawking effect},bookmarksopen=true}{hyperref}
\documentclass[aps,prl,10pt,twocolumn,superscriptaddress]{revtex4}

\usepackage{mathrsfs, amssymb, amsmath, amsfonts, txfonts, latexsym, graphicx, orcidlink}
\usepackage[utf8]{inputenc}

\newcommand{\dd}{\mathrm{d}}
\DeclareMathOperator{\extdm}{d}
\newcommand{\extd}{\extdm \!}

\newcommand{\eq}[2]{\begin{equation} #1 \label{#2} \end{equation}}

\newcommand{\ra}{\rangle}
\newcommand{\la}{\langle}
\newcommand{\T}{\mathcal{T}}
\begin{document}

%%%%%%%%%%%%%%%%%%%%%%
%%% TITLE/ABSTRACT %%%
%%%%%%%%%%%%%%%%%%%%%%

\title{Carroll Hawking effect}

\author{\orcidlink{0000-0003-2093-2377}Ankit Aggarwal}\email{aggarwal@hep.itp.tuwien.ac.at}
\affiliation{Institute for Theoretical Physics, TU Wien, Wiedner Hauptstrasse 8–10/136, A-1040 Vienna,
Austria}

\author{\orcidlink{0000-0002-0449-0081}Florian Ecker}\email{fecker@hep.itp.tuwien.ac.at}
\affiliation{Institute for Theoretical Physics, TU Wien, Wiedner Hauptstrasse 8–10/136, A-1040 Vienna,
Austria}

\author{\orcidlink{0000-0001-7980-5394}Daniel Grumiller}\email{grumil@hep.itp.tuwien.ac.at}
\affiliation{Institute for Theoretical Physics, TU Wien, Wiedner Hauptstrasse 8–10/136, A-1040 Vienna,
Austria}

\author{\orcidlink{0000-0002-2894-4121}Dmitri Vassilevich}\email{dvassil@gmail.com}
\affiliation{CMCC-Universidade Federal do ABC, Avenida dos Estados 5001, CEP 09210-580, Santo Andr\'e, S.P. Brazil}

\date{\today}

\begin{abstract}
Carroll black holes with an associated Carroll temperature were introduced recently. So far, it is unclear if they exhibit a Hawking-like effect. To solve this, we study scalar fields on Carroll black hole backgrounds. Inspired by anomaly methods, we derive a Hawking-like energy-momentum tensor compatible with the Carroll temperature and the Stefan--Boltzmann law. Key steps in our derivation are the finiteness of energy at the Carroll extremal surface and compatibility with the Carroll Ward identities, thereby eliminating, respectively, the Carroll-analogs of the Boulware and Unruh vacua.
\end{abstract}

\maketitle

%%%%%%%%%%%%%%%%%%%%%%
%%% BODY OF LETTER %%%
%%%%%%%%%%%%%%%%%%%%%%

%%% PRL WORD LIMIT: 3750 words between abstract and acknowledgments; each equation line counts 16 words 

\section{Introduction}

Carroll symmetries \cite{Levy1965,SenGupta1966OnAA} were long overlooked by physicists until their omnipresence was recognized. Their ubiquity is partly due to the fact that Minkowski space, crucial for quantum field theories, exhibits a Carroll structure at null infinity. Remarkably, the asymptotic symmetries of asymptotically flat spacetimes known as Bondi, van~der~Burgh, Metzner, and Sachs (BMS) symmetries \cite{Bondi:1962,Sachs:1962} precisely align with conformal Carroll symmetries \cite{Duval:2014uoa,Duval:2014uva,Duval:2014lpa}. Further insights into the Carroll structure at null infinity can be found in Refs.~\cite{Ciambelli:2018wre,Figueroa-OFarrill:2021sxz,Herfray:2021qmp,Mittal:2022ywl,Campoleoni:2023fug}.

Additionally, it was realized that generic null hypersurfaces, prevalent in general relativity, possess a Carroll structure \cite{Penna:2015gza,Penna:2018gfx,Donnay:2019jiz,Ciambelli:2019lap,Redondo-Yuste:2022czg,Freidel:2022vjq,Gray:2022svz,Ciambelli:2023mir,Ciambelli:2023mvj}. Hence, Carrollian symmetries emerge in both pillars of theoretical physics, quantum field theories and general relativity. A prominent application in both contexts is the Carrollian approach to flat space holography, notably in three \cite{Barnich:2006av,Bagchi:2010zz,Bagchi:2012yk,Barnich:2012xq,Bagchi:2012xr,Barnich:2012rz,Bagchi:2013lma,Bagchi:2014iea,Barnich:2015mui,Campoleoni:2015qrh,Bagchi:2015wna,Bagchi:2016bcd,Jiang:2017ecm,Grumiller:2019xna,Apolo:2020bld} and four dimensions \cite{Ciambelli:2018wre,Figueroa-OFarrill:2021sxz,Herfray:2021qmp,Donnay:2022aba,Bagchi:2022emh,Campoleoni:2022wmf,Donnay:2022wvx,Mittal:2022ywl,Campoleoni:2023fug,Bagchi:2023fbj,Saha:2023hsl,Salzer:2023jqv,Nguyen:2023vfz,Nguyen:2023miw,Bagchi:2023cen,Mason:2023mti,Have:2024dff}. 

Carrollian spacetimes are characterized by a Carroll metric $h_{\mu\nu}$ with a degenerate signature $(0,+,\dots,+)$. An illustrative example is the limit of the Minkowski metric where the speed of light vanishes, given by $\extd s^2=\lim_{c\to 0}(-c^2\,\extd t^2+\delta_{ij}\,\extd x^i\extd x^j)=\delta_{ij}\,\extd x^i\extd x^j$. Such spacetimes necessitate a Carroll vector $v^\mu$ lying in the kernel of the Carroll metric, i.e., $v^\mu\,h_{\mu\nu}=0$. In the example, the vector field is $v=v^\mu\partial_\mu=\partial_t$ and the Carroll metric is $h_{\mu\nu}=\delta_{ij}\,\delta^i_\mu\,\delta^j_\nu$.

Whenever we had some global spacetime symmetries in physics, it turned out to be fruitful to make them local. For Poincar\'e symmetries, this leads to Einstein--Cartan theories, including general relativity \cite{Hehl:1976kj}. Conversely, Galilean symmetries yield Newton--Cartan theories \cite{Duval:1984cj,Duval:2009vt,Son:2013rqa,Christensen:2013lma,Bergshoeff:2014uea,Hartong:2015zia,Andringa:2012uz,Bergshoeff:2019pij}. It is therefore natural to gauge the Carroll algebra \cite{Hartong:2015xda} and formulate Carroll gravity theories \cite{Henneaux:1979vn, Bergshoeff:2017btm,Ciambelli:2018ojf,Matulich:2019cdo,Grumiller:2020elf,Gomis:2020wxp,Perez:2021abf,Hansen:2021fxi,deBoer:2021jej,Concha:2021jnn,Figueroa-OFarrill:2022mcy,Campoleoni:2022ebj, Miskovic:2023zfz}.

To advance, detailed examination of Carroll gravity theory is crucial \footnote{%
Even seemingly simple constructions, such as geodesics of test-particles \cite{Ciambelli:2023tzb}, can yield significant deviations from expectations.
}. This is particularly manageable in two dimensions (2d), where Carroll gravity allows for powerful mathematical tools \cite{Grumiller:2019fmp,Gomis:2019nih}. These 2d models can be seen as toy models or as dimensional reductions of higher-dimensional Carroll gravity theories. For example, the Carroll limit of the Schwarzschild black hole aligns with a specific 2d Carroll gravity model \cite{Ecker:2023uwm}.

Recently, it was found that these models can feature Carroll black hole solutions with an associated Carroll temperature \cite{Ecker:2023uwm}. 
The presence of such a temperature raises the question of whether a physical quantum process, akin to the Hawking effect \cite{Hawking:1975vcx}, underlies the temperature of Carroll black holes.

Our Letter demonstrates that indeed there exists a Carroll Hawking effect.

To show this, we introduce a matter scalar field in addition to the 2d geometric variables and consider the consequences of the Ward identities associated with diffeomorphisms, Carroll boosts, and Weyl rescalings. The latter turn out to be anomalous, which we show both from a limiting perspective and in an intrinsically Carrollian way. Our main result is an anomaly-induced expectation value for the energy density \eqref{eq:angelinajolie} that is precisely compatible with the 2d Stefan--Boltzmann law, provided we identify the temperature therein with the Carroll temperature derived classically in \cite{Ecker:2023uwm}.

\section{Matter on Carroll backgrounds}

In the present work, we focus on a massless Carrollian scalar field $\phi$ with conformal coupling \cite{Baiguera:2022lsw,Rivera-Betancour:2022lkc}. We briefly summarize the 2d case to fix the notation for the curved space analogs of electric and magnetic scalar fields introduced in \cite{Henneaux:2021yzg}. The Carroll gravity backgrounds we have in mind are Carroll black hole solutions of magnetic Carroll dilaton gravity \cite{Ecker:2023uwm}, but all results in this section are background-independent.

Starting from the Lorentzian action on a manifold $\mathcal{M}$
\eq{
    I=-\frac{1}{2}\int _{\mathcal{M}}\dd ^2x \sqrt{-g}\; g^{\mu \nu}(\partial_\mu\phi)\,(\partial _\nu \phi) 
}{eq:lalapetz}
we introduce pre-ultralocal variables \cite{Hansen:2021fxi} by $V^\mu T_\mu =-1$, $T_\mu E^\mu =0$, $V^\mu E_\mu =0$ as well as $E^\mu E_\nu =\delta ^\mu _\nu +V^\mu T_\nu$ such that the metric is given by $g_{\mu \nu }=-c^2T_\mu T_\nu +E_\mu E_\nu$ and the Lorentzian volume form is $cT\wedge E$. For the Carrollian limit the frame fields are expanded in powers of $c^2$ as $ V^\mu =v^\mu +\mathcal{O}(c^2)$, $T_\mu =\tau _\mu +\mathcal{O}(c^2)$, $E_\mu =e_\mu +\mathcal{O}(c^2)$ and $E^\mu =e^\mu +\mathcal{O}(c^2)$. Local Carroll boosts parametrized by $\lambda (x)$ act as 
\begin{align}\label{eq:Carr_boost}
    \delta _\lambda e&=0 & \delta _\lambda \tau &=-\lambda e  & \delta _\lambda v^\mu &=0 & \delta _\lambda e^\mu &=-\lambda v^\mu ~.
\end{align}
Local Weyl rescalings \cite{Baiguera:2022lsw} parametrized by $\rho (x)$ act on the frame fields as 
\begin{align}\label{Weyl}
    \delta _\rho e&=\rho e & \delta _\rho \tau &=\rho \tau & \delta _\rho v^\mu &=-\rho v^\mu & \delta _\rho e^\mu &=-\rho e^\mu ~.
\end{align}

Switching to a Hamiltonian formulation by defining the pre-ultralocal momentum $\Pi =\frac{c}{\sqrt{-g}}\frac{\delta L}{\delta (V^\mu \partial _\mu \phi )}=\pi +\mathcal{O}(c^2)$ and inserting the pre-ultralocal variables into the action \eqref{eq:lalapetz} yields
\begin{align}
    I=\int_{\mathcal{M}}\!T\wedge E \,\Big(\Pi \,V^\mu \partial _\mu \phi -\frac{1}{2}\Pi ^2-\frac{c^2}{2}(E^\mu \partial _\mu \phi )^2\Big) ~.\label{eq:Ham_action_L}
\end{align}
This is the starting point for obtaining two possible actions for a Carroll invariant scalar field \cite{Henneaux:2021yzg}, which we discuss now. 

\paragraph{Timelike (electric) scalar field.} The electric contraction is obtained by directly sending $c\to 0$ in \eqref{eq:Ham_action_L}, replacing all fields by their leading order expressions, and integrating out the leading order momentum $\pi$,
\eq{
I_{\mathrm{\tiny el}}[\phi ]:=\frac{1}{2}\int_{\mathcal{M}}\!\tau \wedge e \, \big(v^\mu \partial _\mu \phi )^2\,.
}{eq:electric action} 
The spatial dependence of the field $\phi$ is unconstrained, representing the ultralocal character of Carrollian theories. It is straightforward to check that this action is invariant under local Carroll boosts as well as diffeomorphisms, as required. Additionally, the action \eqref{eq:electric action} is invariant under Weyl-rescalings \eqref{Weyl} of the background. 

\paragraph{Spacelike (magnetic) scalar field.} There is a second possibility to contract the Hamiltonian action where the fields are rescaled as $\Pi{\to}c\Pi$, $\phi\to\frac{1}{c}\phi$. Crucially, this rescaling preserves the symplectic form $\delta\Pi\wedge\delta\phi$ on field space. The leading order action
\eq{
I_{\mathrm{\tiny mag}}[\phi ,\pi ]:=\int_{\mathcal{M}}\!\tau \wedge e\, \Big(\pi v^\mu \partial_\mu \phi -\frac{1}{2}(e^\mu \partial_\mu \phi )^2\Big)
}{eq:magnetic action}
does not permit integrating out the momentum $\pi$ since its quadratic term cancels in the contraction. Instead, $\pi$ acts as a Lagrange multiplier enforcing time-independence of the scalar field. Under local Carroll boosts, the momentum transforms as $\delta_\lambda\pi=-{\lambda}e^\mu\partial_\mu\phi$ such that the total action is invariant. 
This transformation is compatible with Weyl-rescalings \eqref{Weyl} if they act on $\pi$ as $\delta_\rho\pi=-\rho\,\pi$, rendering the magnetic action \eqref{eq:magnetic action} Weyl-invariant as well. 

\paragraph{Carroll Ward identities.} Both examples of classical matter actions are invariant under local Carroll boosts and local Weyl rescalings, which leads to Ward identities for the associated Carroll energy-momentum tensor (CET). Taking the electric scalar as an example, we define the one-forms $T^{(v)}$ and $T^{(e)}$ by 
\begin{align}\label{eq:stress_var}
    \delta I_{\mathrm{\tiny el}}=-\int _{\mathcal{M}}\tau \wedge e \, \Big(T^{(v)}_\mu \delta v^\mu +T^{(e)}_\mu \delta e^\mu \Big)
\end{align}
which implies that their components transform under Carroll boosts as
\begin{align}
    \delta _\lambda T^{(v)}_\mu &=\lambda T^{(e)}_\mu & \delta _\lambda T^{(e)}_\mu &=0 ~.
\end{align}
The CET 
\begin{align}\label{eq:rank2stress}
    T^\mu {}_\nu =v^\mu T^{(v)}_\nu +e^\mu T^{(e)}_\nu 
\end{align}
is gauge invariant \cite{Hartong:2014oma,Hartong:2014pma,Baiguera:2022lsw}. Contracting the arbitrary variation \eqref{eq:stress_var} with a Carroll boost \eqref{eq:Carr_boost} yields the Carroll boost Ward identity
\begin{align}\label{eq:boost_ward}
    T^{(e)}_\mu v^\mu &=e_\mu T^\mu {}_\nu v^\nu =0
\end{align}
while contracting with an infinitesimal diffeomorphism yields
\begin{align}\label{eq:diff_ward_id}
    \frac{1}{e}\partial _\mu \big(eT^{(v)}_\nu v^\mu +eT^{(e)}_\nu e^\mu \big)=-T^{(v)}_\mu \partial _\nu v^\mu -T^{(e)}_\mu \partial _\nu e^\mu 
\end{align}
where $e:=\det(\tau _\mu,e_\mu)$. Weyl-invariance additionally requires the trace of the CET to vanish,
\begin{align}
    T^\mu {}_\mu =v^\mu T^{(v)}_\mu +e^\mu T^{(e)}_\mu =0 ~.
\end{align}
As we shall prove in our Letter, this last Ward identity becomes anomalous in the quantum theory.

\paragraph{Carroll--Schwarzschild black hole.} Our prototypical example for a Carroll black hole background is the spherically reduced Carroll--Schwarzschild spacetime \cite{Hansen:2021fxi,Perez:2021abf,Ecker:2023uwm} 
\begin{align}\label{eq:Carr_Schwarzschild}
    \tau &=\sqrt{\xi} \,\dd \Tilde{t} &  e&=\frac{\dd r}{\sqrt{\xi} } & v&=-\frac{1}{\sqrt{\xi} } \partial _{\Tilde{t}} & \xi &=1-\frac{r_s}{r}
\end{align}
where we used temporal and radial coordinates $(\Tilde{t},r)\in\mathbb{R}\times(r_s,\infty)$. We decorated the time coordinate with a tilde since later we shall use $t$ for the Wick-rotated time. While the full solution of 2d Carroll dilaton gravity also contains the dilaton, we do not display it here since the matter theories we consider do not couple to it. The locus $r=r_s$ represents the Carroll extremal surface of this geometry \cite{Ecker:2023uwm}.

\section{Carroll Hawking effect as a limit}

In this section, we extend to the Carrollian case the method of Christensen and Fulling \cite{Christensen:1977jc} that allows to recover the expectation values of the full Lorentzian energy-momentum tensor through the conformal anomaly. We do so by carefully implementing the Carrollian limit together with the definition of the semi-classical theory. 

Our starting point is the classically Weyl invariant ``electromagnetic'' scalar action \cite{Ciambelli:2023xqk}
\begin{align}\label{eq:electromag_action}
    I_{\rm em}=\int _\mathcal{M}\!\frac{\tau \wedge e}{\sqrt{g_1g_2}}\ \left( g_1\, (v^\mu \partial_\mu \phi)^2 + g_2\, e^\mu e^\nu (\partial_\mu \phi)(\partial_\nu \phi) \right)
\end{align}
which has the terms from both electric and magnetic actions \eqref{eq:electric action}, \eqref{eq:magnetic action} with coupling constants $g_1$ and $g_2$. This action is not manifestly invariant under local Carroll boosts, but we will remedy this by taking appropriate limits of $g_1$ and $g_2$ \footnote{The action \eqref{eq:electromag_action} can be brought into a manifestly Carroll boost-invariant form by adding to it a term that vanishes on all Carroll black hole backgrounds \cite{Ecker:2024}, so there is no issue with Carroll boost-invariance.}. We rewrite the action \eqref{eq:electromag_action} more suggestively as
\begin{equation}\label{eq:action_fid_metric}
    I_{\rm em}=\int _\mathcal{M}\dd ^2x \sqrt{G}\,\Big(G^{\mu \nu }\partial _\mu \phi \partial _\nu \phi \Big)
\end{equation}
where we introduced a fiducial metric 
\begin{align}\label{eq:fiducial_metric}
    G_{\mu \nu }(g_1,g_2)=\frac{1}{g_1}\tau_\mu \tau _\nu +\frac{1}{g_2}e_\mu  e_\nu ~.
\end{align}
The limit $g_1\to \infty $, $g_2=1$ renders $G_{\mu \nu }$ Carrollian. Comparing with \eqref{eq:magnetic action}, one can show that this limit corresponds to a magnetic limit on the level of the scalar action \cite{deBoer:2023fnj}. The inverse of the fiducial metric \eqref{eq:fiducial_metric} is $G^{\mu\nu}(g_1,g_2)=g_1{v^\mu}v^\nu+g_2e^{\mu}e^{\nu}$. 

From this point we formally treat the electromagnetic scalar theory as a Euclidean theory, which makes it natural to define the path integral measure by
\begin{align}
    1=\int \mathcal{D}\phi \exp \left( -\int_{\mathcal{M}}\, \dd^{2}x \sqrt{G(g_1,g_2)} \, \phi^2 \right)\,.\label{meas}
\end{align}
This definition is invariant under diffeomorphisms as well as local Carroll boosts for arbitrary $g_1$, $g_2$ but breaks Weyl symmetry. Non-invariance of the path integral measure under a classical symmetry of the action is the hallmark of anomalies \cite{Fujikawa:1979ay}, so we expect a Weyl anomaly and confirm this expectation below.

The partition function with the measure \eqref{meas},
\begin{align}
    Z=\int \mathcal{D}\phi \exp \bigg(-\int _{\mathcal{M}}\dd ^2 x\sqrt{G}\,\phi A \phi \bigg) =(\det\,A)^{-\frac{1}{2}}
\end{align}
is given in terms of the determinant of the Laplace-type operator $A=-G^{\mu\nu}\nabla_\mu\nabla_\nu$, where $\nabla$ is the Levi--Civit\'a connection associated with $G_{\mu\nu}$. The broken Weyl symmetry implies that the effective action $W=-\ln{Z}$ is not invariant under rescalings. The associated trace anomaly is the standard result \cite{Christensen:1977jc,Grumiller:2002nm},
\eq{
    \langle \mathcal T^{\mu }{}_\mu \rangle =\frac{1}{24\pi}\,R^{(G)} 
}{eq:anomalous_trace}
where we used $\delta_{\rho}G_{\mu\nu}=2\rho\,G_{\mu\nu}$ and $R^{(G)}$ is the Ricci scalar associated with $\nabla$. The expectation value of the fiducial energy-momentum tensor is defined by $\delta{W}=\frac{1}{2}\int_{\mathcal{M}}\dd^2x\sqrt{-G}\,\langle\mathcal{T}_{\mu\nu}\rangle\delta{G}^{\mu\nu}$. 

Let us consider now the Carroll--Schwarzschild background \eqref{eq:Carr_Schwarzschild}. In the Ricci scalar $R^{(G)}=\frac{2g_2r_s}{r^3}$ the parameter $g_1$ drops out because it can be absorbed into a redefinition of time $\Tilde{t}$. At this stage, the components $\langle\mathcal{T}^\mu{}_\nu\rangle$ are not Carrollian as they still depend on the $g_i$ and thus do not satisfy the Carroll boost Ward identity. However, in addition to \eqref{eq:anomalous_trace} they satisfy the Euclidean diffeomorphism Ward identities, $\nabla^\mu\langle\mathcal{T}_{\mu\nu}\rangle=0$, which can be solved up to two integration constants in the static case. Pretending that $G_{\mu\nu}$ describes a Wick-rotated Lorentzian geometry, we undo this Wick-rotation, $\Tilde{t}\to{it}$, $v\to{iv}$, $\tau\to-i\tau$ and define adapted null coordinates \footnote{The definitions \eqref{eq:lightcone_coords} in principle allow additional shifts $\Tilde{t}\to\alpha\Tilde{t}$, $z\to\alpha\sqrt{\frac{g_1}{g_2}}z$ with some $\alpha(g1,g_2)$ which, however, do no affect our result.}
\eq{
    x^\pm =\frac{1}{\sqrt{2}}\Big(\sqrt{g_2 }\, t \pm \sqrt{g_1 }z\Big) \qquad\qquad \frac{\dd z}{\dd r}=\frac{1}{1-\frac{r_s}{r}}
}{eq:lightcone_coords}
in terms of which the fiducial metric is
\eq{
    \dd s^2_{(G)}=-\frac{2}{g_1 g_2 }e^{2\omega }\dd x^+ \dd x^- \qquad\qquad \omega =\frac{1}{2}\ln \Big(1-\frac{r_s}{r}\Big) ~.
}{eq:lorentz_metric}
We solve the Ward identities by
\begin{align}\label{eq:flux_comp}
    \la \T _{\pm \pm }\ra =\frac{1}{24\pi g_1 }\Big(\partial _z^2\omega -(\partial _z \omega )^2\Big) +\frac{t_\pm}{g_1 } &&  t_\pm\in \mathbb{R} 
\end{align}
where $t_\pm$ are constants of integration. A locally Carroll boost-invariant CET is only defined as a $(1,1)$-tensor in a static coordinate system \cite{Baiguera:2022lsw}. Therefore, we invert the transformation \eqref{eq:lightcone_coords} and pull up one index with $G^{\mu \nu}$, leading to 
\begin{subequations}
\begin{align}
     \la \T^{t} {}_{t} \ra &=-\frac{g_2}{24\pi }\Big(\partial _r^2\xi -\frac{1}{4\xi }(\partial _r\xi )^2+12\pi \frac{t_++t_-}{\xi }\Big)\\
    \la \T^{t} {}_r\ra &=-\frac{\sqrt{g_1g_2}}{2}\frac{t_+-t_-}{\xi ^2}\\
    \la \T^r{}_{t} \ra &=\frac{g_2}{2}\sqrt{\frac{g_2}{g_1}}(t_+-t_-) \\
    \la \T^r{}_r\ra &=\frac{g_2}{24\pi }\Big(-\frac{(\partial _r\xi )^2}{4\xi }+12\pi \frac{t_++t_-}{\xi }\Big)
\end{align}
\end{subequations}
with $\xi=1-\frac{r_s}{r}$. The flux components can then be expressed as
\begin{align}\label{eq:flux_comps}
    \la \T_{\pm \pm }\ra =\frac{\xi }{2g_2}\la \T^r{}_r\ra \pm \frac{t_+-t_-}{2}-\frac{\xi }{2g_2}\la \T^{t} {}_{t} \ra ~.
\end{align} 

\paragraph{Magnetic limit.} One way to obtain a local Carroll boost-invariant theory is to set $g_2=1$ and $g_1\to\infty$ corresponding to a magnetic contraction. In this limit, $\la\T^{t}{}_r\ra\rightarrow\infty$ unless we assume $t_+-t_-=\frac{t_0 }{\sqrt{g_1}}$ with some fixed constant $t_0$. With this assumption, we obtain in the magnetic limit
\begin{subequations}
\begin{align} \label{eq:T limit sch}
    \la \T^{t} {}_{t} \ra \to \la T^{t} {}_{t} \ra &=-\frac{1}{24\pi }\Big(\partial _r^2\xi -\frac{(\partial _r\xi )^2}{4\xi }+ \frac{24\pi t_+}{\xi }\Big)\\
    \la \T^{t} {}_r \ra \to \la T^{t} {}_r\ra &=-\frac{1}{2}\frac{t_0 }{\xi ^2}\\
    \la \T ^r{}_{t} \ra \to \la T^r{}_{t} \ra &=0\\
    \la \T^r{}_r\ra \to \la T^r{}_r\ra &=\frac{1}{24\pi }\Big(-\frac{(\partial _r\xi )^2}{4\xi }+ \frac{24\pi t_+}{\xi }\Big)~.
\end{align}
\end{subequations}
This result satisfies the Carroll boost and diffeomorphism Ward identities \eqref{eq:boost_ward}, \eqref{eq:diff_ward_id} for a CET where
 \begin{align}
     \la T^{(v)}_\nu \ra =-\la T^\mu {}_\nu \ra \tau _\mu && \la T^{(e)}_\nu \ra =\la T^\mu {}_\nu \ra e_\mu~. 
\end{align}
The trace Ward identity stays anomalous after the limit. Taking the limit of the flux components \eqref{eq:flux_comps} leads to
\begin{align}\label{eq:flux_limit}
    \la T_{\pm \pm }\ra  := \lim _{g_1\to \infty } \la \T_{\pm \pm }\ra \Big \vert _{g_2=1}=\frac{1 }{96\pi }\Big(2\xi \partial _r^2\xi -(\partial _r\xi )^2\Big)+t_+ 
\end{align}
which shows that both fluxes have to agree, $\la T_{++}\ra =\la T_{--}\ra $. This is unlike the situation in a true Lorentzian theory, where $\la\T_{\pm\pm}\ra$ would be associated with in- and outgoing matter fluxes. They would behave independently from each other, according to the physical situation at hand. The fact that both fluxes have to agree in the present case is just another manifestation of no energy flux being possible in a Carrollian theory \cite{deBoer:2021jej}. It furthermore implies that not all vacuum choices of the analogous Lorentzian theory are possible anymore. In particular, \emph{local Carroll boost-invariance is inconsistent with the Unruh vacuum}. The Boulware vacuum is ruled out by demanding finite energy density 
\begin{align}
    \langle \mathcal{E} \rangle =-\tau_\mu \la T^\mu {}_\nu \ra v^\nu =\frac{1}{24\pi }\Big(\partial _r^2\xi -\frac{(\partial _r\xi )^2}{4\xi }\Big)+ \frac{t_+}{\xi } 
\end{align}
at the Carroll extremal surface. This leads to the unique choice $t_+=\frac{1}{96\pi{r}_s^2}$ and defines the Carroll analog of the Hartle--Hawking vacuum with asymptotic energy density
\eq{
    \lim_{r\to\infty} \langle \mathcal{E}_{\mathrm{\tiny HH}}\rangle =\frac{1}{96\pi r_s^2}=\frac{\pi}{6}\,T^2 ~.
}{eq:angelinajolie}
In the second equality we used the result for the Carroll temperature $T$ of the Carroll--Schwarzschild background, $T^{-1}=4\pi{r}_s$. This equality is our main result and shows that the asymptotic energy density \eqref{eq:angelinajolie} is compatible with the 2d Stefan-Boltzmann law. %, $\mathcal{E}_{\mathrm{\tiny HH}}=\frac\pi6T^2$.

\section{Conformal anomaly in Carrollian theories} 

Instead of taking Carrollian limits, we consider in this section the magnetic scalar action \eqref{eq:magnetic action} from the start. Plugging it into the path integral yields
\begin{align}
    Z=\int \mathcal{D}\pi \mathcal{D}\phi \exp \Big(-I_{\mathrm{\tiny mag}}[\phi ,\pi ]\Big) ~.\label{Zmag1}
\end{align}
Integrating out $\pi $ produces a functional $\delta$-function $\delta(v^\mu\partial_\mu\phi)$ so that we remain with a path integral over a 1d time-independent scalar field, but with a Jacobian factor $\mathcal{J}=(\det(v^\mu\partial_\mu))^{-1}$. The operator $v^\mu\partial_\mu$ contains a derivative along the time direction but no derivative along the spatial direction. This means that the operator is not elliptic, so there is no regular method known to us to define its determinant \footnote{This situation is similar to one with a Faddeev--Popov determinant in axial gauge on a curved background \cite{Vassilevich:1995cz}.}. Since a direct method fails, we try a less direct one. 

We assume that the path integral \eqref{Zmag1} exists and write a conformal variation of the corresponding effective action [see \eqref{Weyl}] 
\eq{
    \delta _\rho W
    =-\int _{\mathcal{M}}\tau \wedge e\,\big(\la T^{(v)}_\mu \ra v^\mu +\la T^{(e)}_\mu \ra  e^\mu \big) \, \rho ~.
}{eq:var_effact}
Demanding that the conformal anomaly is local and Carroll boost-invariant, the only choice with the correct mass dimension,
\eq{
   \delta _\rho W= -\alpha_1 \int _{\mathcal{M}}\dd ^2x \det (\tau ,e) \,R\, \rho 
}{ca-mag}
contains an undetermined constant $\alpha_1$ that we shall fix below. Here, $R$ is the Carroll boost-invariant Carrollian curvature scalar \cite{Ecker:2023uwm} given in terms of the 2d Carroll boost connection $\omega$ by $2\dd\omega=R\,\tau\wedge{e}$. For the Carroll--Schwarzschild background \eqref{eq:Carr_Schwarzschild}, it reads
\eq{
    R=-\partial _r^2\xi -2\partial _r\frac{\xi }{r}=\frac{2r-2r_s}{r^3}~.
}{eq:whatever}

The Ward identities for Carroll boosts \eqref{eq:boost_ward} and diffeomorphisms \eqref{eq:diff_ward_id} read in this gauge
\begin{align}
    \la T^{(e)}_t\ra &=0\\
     \partial _r \la T^{(e)}_r\ra +\frac{\partial _r\xi }{\xi }\la T^{(e)}_r\ra &=-\frac{\partial _r \xi }{2\xi ^2} \la T^{(v)}_t\ra ~.
\end{align}
Together with the anomalous trace given by \eqref{eq:var_effact}, \eqref{ca-mag} they have a family of exact solutions ($a\in\mathbb{R}$)
\begin{align}
    \la T^{(e)}\ra &=\Big[\frac{\alpha _1}{\xi ^{\frac{3}{2}}}\Big(\frac{r_s^2}{4r^4}-\frac{r_s}{3r^3}\Big)+\frac{a}{\xi ^\frac{3}{2}}\Big]\,\dd r\\
     \la T^{(v)}\ra &=\Big[\la T^{(e)}_r \ra \xi -\sqrt{\xi }\alpha _1 R\Big]\, \dd t+\la T^{(v)}_r\rangle \,\dd r\,.
\end{align} 
The component $\la T^{(v)}_r\rangle $ remains undetermined. This happens since, on static backgrounds, the Ward identity \eqref{eq:diff_ward_id} for $\nu=t$ is satisfied automatically. Thus, in contrast to the Lorentzian case, we do not have enough conditions to define all components of the CET.

The energy density $\langle\mathcal{E}\rangle=\la{T}^{(v)}_\mu \ra{v}^\mu$ is finite at $r\to{r}_s$ if we choose the integration constant $a=\frac{\alpha_1}{12r_s^2}$, producing an asymptotic energy density
\eq{
    \lim _{r\to \infty }\langle \mathcal{E}\rangle  =\frac{\alpha _1}{12r_s^2} 
}{eq:result}
which coincides precisely with the Carroll--Hartle--Hawking energy density \eqref{eq:angelinajolie} for $\alpha_1=\frac{1}{8\pi}$. 

\section{Conclusion}
We have shown that the semi-classical theory of a free scalar field on a Carroll black hole background exhibits a Carroll analogue of the Hawking effect. It manifests through a non-vanishing energy density in the asymptotic region compatible with the Stefan--Boltzmann law \eqref{eq:angelinajolie}. However, as a consequence of the Ward identities the energy flux in any Carrollian field theory has to vanish which prevents the Carroll black hole from evaporating. This implies that the Unruh vacuum is incompatible with Carroll symmetries, leaving only the Carroll analogue of the Hartle--Hawking vacuum as a viable semi-classical vacuum state. For proving this we used anomaly-based arguments going back to Christensen and Fulling. The Carrollian quantum theory is thereby defined by first regularizing the classical action \eqref{eq:electromag_action} and then quantizing, removing the regulator only in the end. While this initially breaks local Carroll boost invariance we justify the procedure by the absence of a Carroll boost anomaly after removing the regulator.

We conclude by mentioning a number of further directions. The derivation of the Carroll Hawking effect in this work did not rely on the specific form of the scalar field action but rather solved for the vacuum expectation values of the energy-momentum tensor using symmetry-based arguments. It would be interesting to see if the same conclusion can be reached by following a microscopic derivation along the lines of Hawking's original work \cite{Hawking:1975vcx}. Another possibly interesting problem would be to consider the backreaction of matter on the Carroll black hole backgrounds, classically as well as semi-classically.

%%% WORD COUNT TOLL HERE: 2477 + 16*44 = 3181 words; the 44 comes from the 44 lines of eqs.; 

%\acknowledgments
\bigskip

\paragraph{Acknowledgements} 
We thank Jelle Hartong, Alfredo P\'erez, Stefan Prohazka, and Ricardo Troncoso for discussions on Carroll black holes.

This work was supported by the Austrian Science Fund (FWF), projects P 33789, and P 36619, by the S\~ao Paulo Research Foundation (FAPESP), project 2021/10128-0, and by the National Council for Scientific and Technological Development (CNPq), project 304758/2022-1. AA, FE, and DG acknowledge support by the OeAD travel grant IN 04/2022 and thank Rudranil Basu for hosting them at BITS Pilani in Goa in February 2024 through the grant DST/IC/Austria/P-9/202 (G).

%%%%%%%%%%%%%%%%%%
%%%            %%%
%%% REFERENCES %%%
%%%            %%%
%%%%%%%%%%%%%%%%%%

%\bibliographystyle{fullsort}

\end{document}